\documentclass[epj]{svjour}
\usepackage{graphicx}
\begin{document}
\title{On the existence of stationary states during granular compaction}
\author{Philippe Ribi\`ere\inst{1}, Patrick Richard\inst{1}, Pierre Philippe\inst{2}, Daniel Bideau\inst{1} and Renaud Delannay\inst{1}}
\institute{Groupe Mati\`ere Condens\'ee et Mat\'eriaux, UMR CNRS 6626, Universit\'e de Rennes 1, Campus de Beaulieu, F-35042 Rennes cedex, France
\and Cemagref, Aix-en-Provence, Le Tholonet BP 31, F-13612 Aix-en-Provence Cedex 1, France}
\date{}
\abstract{
When submitted to gentle mechanical taps a granular packing
slowly compacts until 
it reaches a stationary state that depends 
on the tap characteristics.
The properties of such stationary states are
experimentally investigated. 
The influence of the initial state, taps properties and
tapping protocol are studied. 
The compactivity of the packings is determinated.
Our results strongly
support the idea that the stationary states are genuine 
thermodynamic states.  
\PACS{ 
{45.70.-n}{Granular systems.} \and 
{45.70.Cc}{Static sandpiles; granular compaction} \and
{81.05.Rm}{Porous materials; granular materials} 
}
}
\authorrunning{P. Ribi\`ere et al}
\titlerunning{Experimental compaction of anisotropic granular media}
\maketitle
\section{Introduction}\label{sec:intro}
In the absence of an external drive, granular materials rapidly come 
to rest. This is a consequence of their dissipative interactions
and of the irrelevance of thermal energy (in this case thermal energy is
negligible compared to the energy needed to move a grain). 
Thus standard thermodynamics is not applicable to those systems.
Edwards and Oakeshott~\cite{Edwards1989a}
proposed a non-standard thermodynamic description 
for static granular media. This theory
postulates that a granular packing at rest can be described
by suitable ensemble averages over
its blocked ``jammed'' states. If one assumes that 
all the available blocked states have the same probability
to occur and that the volume is analog to the energy 
of thermal systems, the configurational
entropy of a granular packing is $S=\lambda\ln\Omega(V,N)$,
 where $\lambda$ is the
 equivalent of the Boltzmann constant and $\Omega(V,N)$ the number
 of mechanically stable configurations of $N$ particles
 in the volume $V$.
A
temperature-like state variable, the compactivity
$X=\partial V/\partial S$ can then be introduced. 
This theory was partially investigated by 
recent experiments~\cite{Knight1995,Philippe2002,Philippe2003,Danna2003c}, 
numerical simulations of models~\cite{Makse2002a,Barrat2000b,Berg2002,Depken2005,Fierro2002a}.
These experiments have established that a granular system subjected to
a tapping dynamics slowly compacts (i.e. the packing fraction increases)
and reaches a
stationary state that depends on the tapping
intensity. The existence of a stationary state is
the  first essential step to justify the 
validity of a thermodynamic-like description for granular packings.\\
Here we  experimentally study the dynamics of granular
media submitted to gentle mechanical taps.
We mainly focus on the properties
of the stationary states and discuss the validity of
such configurations as genuine thermodynamic states. Using the fluctuations
of the packing fraction at stationary state we will 
determine the compactivity introduced by Edwards and Oakeshott~\cite{Edwards1989a}
and compare our results  with those obtained for granular
packings submitted to fluid flow pulses~\cite{Schroter2005}.\\
The outline of this paper is the following. We first describe the 
experimental setup. Then, in section~\ref{sec:recall},
we rapidly recall the previous
results obtained on the relaxation of the packing fraction 
during compaction. 
Section~\ref{sec:compaction_decompaction} is devoted 
to the influence of the initial conditions on the stationary state
and section~\ref{sec:shape_tap} reports the dependence of our results
on the frequency used to generate the taps. 
The cases of annealing and memory effects are studied in 
section~\ref{sec:aging}. Finally, section~\ref{sec:conclu} is
devoted to the conclusions of this work.
\section{Experimental setup}\label{sec:setup}
\begin{figure}[htbp]
\includegraphics*[width=0.5\textwidth]{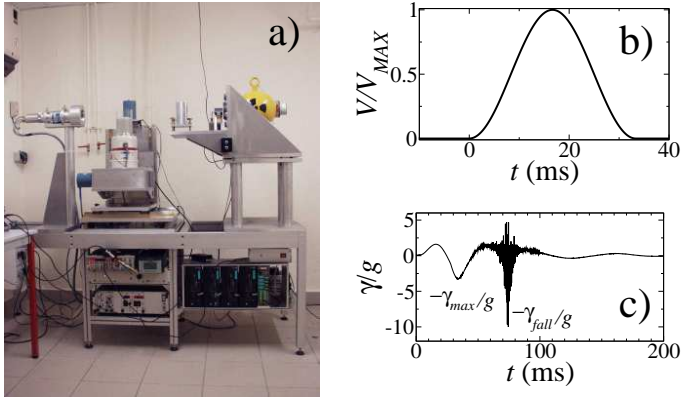}
\caption{a) Picture of the experimental setup, b) the
signal controlling the tap (here $f=30\mbox{ Hz}$) and c) the
acceleration monitored by the accelerometer (here $\Gamma\approx3.3$).}\label{fig:sketch}
\end{figure}

The experiments (Fig.~\ref{fig:sketch}a) are performed with
$d=1\mbox{ mm}$ diameter glass spheres placed in
a glass cylinder of diameter $D=10\mbox{ cm}$.
The cylinder containing the grains is tapped
vertically at regular intervals ($\Delta t = 1\mbox{ s}$).
Each tap is controlled by an entire cycle of a sine wave at a fixed
frequency $f=30~\mbox{Hz}$: $V(t)=V_{MAX}(1-\cos{(2\pi f t))/2}$ for
$0<t<1/f$ and $V(t)=0$ elsewhere (Fig.~\ref{fig:sketch}b).
This applied voltage is connected
to an electromagnetic
exciter (LDS V404) which induces a vertical
displacement to a moving part supporting the container and the beads.
The resulting motion of the whole system is monitored by an
accelerometer at the bottom of the container. This motion
is more complicated than a simple sine wave: at first the 
system undergoes a positive acceleration followed by a 
negative acceleration with a minimum equal to $-\gamma_{max}$.
After the applied voltage stops, the system relaxes to its
normal position (Fig.~\ref{fig:sketch}c).
When $\gamma_{max}$ is large enough, 
the bead packing takes off from the bottom of the container
and achieves a flight until it crashes back to the bottom.
This crash is visible on the signal of the accelerometer:
a negative acceleration $-\gamma_{fall}$
corresponding to the fall of the
grains is clearly visible.
The control parameters
are the frequency $f$ used to generate the tap and the tapping intensity $\Gamma=\gamma_{max}/g$, where
$g=9.81\mbox{ m.s}^{-2}$.
The packing fraction is measured using a $\gamma$-ray absorption
setup~\cite{Philippe2002}.
The measure is deduced from the transmission ratio of the horizontal
collimated $\gamma$ beam through the packing: $T=A/A_0$, where $A$ and $A_0$
are, respectively, the activities counted  on the detector with and without 
the presence of the granular medium. From Beer-Lambert law for absorption,
we can derive an estimation of the volume fraction in the probe
zone: $\rho\approx-(\mu D)^{-1}\ln(T)$, where $\mu$ is the absorption
coefficient of the beads. It was evaluated experimentally to
$\mu\approx 0.188\mbox{ cm}^{-1}$ for our $\gamma$
beam of energy $662\mbox{ keV}$ ($^{137}$Cs source).
The collimated $\gamma$ beam is nearly cylindrical with a diameter of $10\mbox{ mm}$ and intercepts perpendicularly the vertical axis of the vessel.
In order to reduce the relative fluctuations due to number of emitted photons
we use an acquisition time of $300$ seconds for each measure
which
leads to a precision of $0.001$.
The packing fraction of the sample is estimated from the ratio $T$
averaged on approximately $7$ cm height from the bottom of the cylinder.
\section{Packing fraction relaxation laws}\label{sec:recall}
The first quantity of interest in compaction is the packing fraction 
(or density) $\rho$, defined as the ratio of the volume of the grains
to the total volume occupied by the packing. 
A few characteristic values of $\rho$ for mono-sized 
sphere packings have to be reminded. The maximal packing fraction
reached in a random packing of spheres (the so-called random close
packing fraction) is $\rho_{RCP}\approx0.64$. This value is significantly 
lower than the maximal packing fraction
obtained for face-centered-cubic (or hexagonal compact) packing
($\rho_{max}\approx0.74$). Another limit is the so-called random loose
packing corresponding to the less dense  mechanically stable packing  
($\rho_{RLP}\approx 0.58$).

\begin{figure}
\begin{center}
\includegraphics*[width=0.5\textwidth]{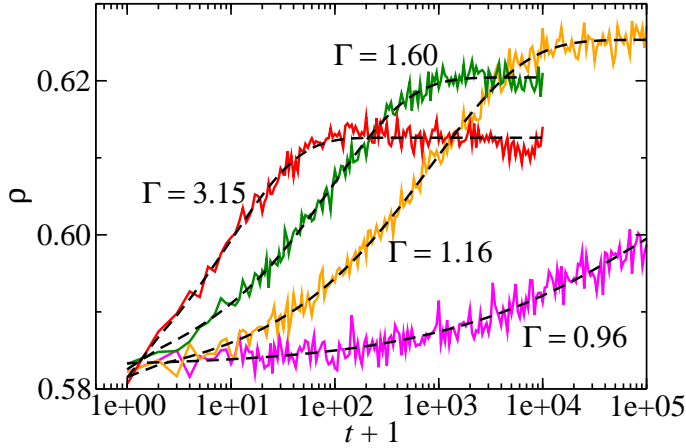}
\caption{Packing fraction versus the number of taps in our
experiments. The curves are KWW fits (see text). The tap frequency is
$f=30\mbox{ Hz}$.}\label{fig:compa}
\end{center}
\end{figure}

\begin{figure}[htbp]
\includegraphics*[width=0.5\textwidth]{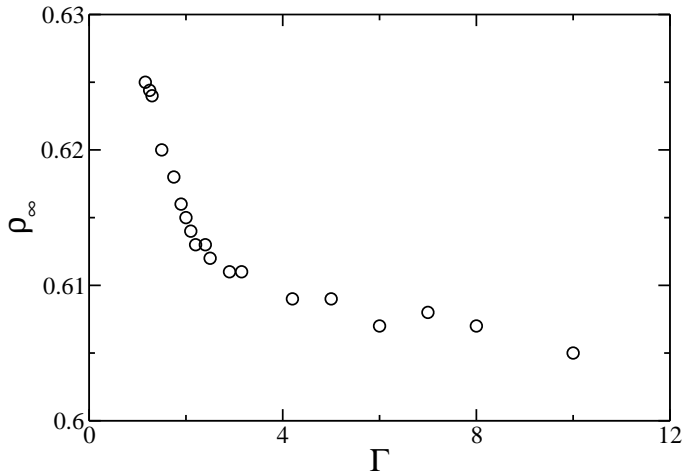}
\caption{Packing fraction of the stationary state versus the tapping intensity
for $f=30\mbox{ Hz}$}\label{fig:compa_finale}
\end{figure}

In a pioneering paper, Knight et al.~\cite{Knight1995} in Chicago first 
considered packing fraction  relaxation law in granular compaction.
They showed that, starting from a loose packing of beads confined 
in a very narrow and tall tube (diameter $1.88$~cm and $87$ cm high, 
bead diameter 2 mm),
a succession of vertical taps 
induces a progressive and very slow compaction of the system. 
They found that the relaxation law can be
well fitted by an inverse-logarithmic law, the so-called Chicago fit: 
$$\rho(t)=\rho_{\infty} - \frac{\rho_0-\rho_{\infty}}{1+B \ln\left(1 + {t}/{\tau} \right)}.$$
For a given frequency $f$,
 the fitting parameters $\rho_{\infty}$,
$\rho_0$, $B$ and $\tau$ essentially depend on $\Gamma$.
The small number of grains in a tube diameter ($\approx$ 10) 
allows for a local
measurement of the packing fraction
with a capacitive
method and prevents any convection in the packing.
Nevertheless it induces strong boundary effects that may be
responsible for crystallization (some packing fraction
values obtained are well above the random close packing limit)
visible in some
of the experiments reported in~\cite{Nowak1997a}
and in the values of $\rho_\infty$ obtained in~\cite{Knight1995}.
\\
More recently Philippe and Bideau~\cite{Philippe2002}
carried out new compaction experiments 
using the setup described in section~\ref{sec:setup}
with about 100 grains in the tube diameter.
This restricts the boundary effects but, 
contrary to the Chicago group's experiments, allows convection.
 The relaxation law obtained by these
authors differs significantly from
those obtained by Knight et al.~\cite{Knight1995},
especially for the long-time behavior. Indeed whereas in
previous experiments no clear 
evidence of convergence to a steady state has been established, 
such an evidence in definitely produced by our experiments
and may
correspond to a dynamical balance between convection and compaction.
The relaxation is better fitted by 
the Kohlrausch Williams Watts law (KWW law) - a 
stretched exponential - :
$$\rho(t)=\rho_{\infty}-\left(\rho_{\infty}-\rho_0\right)\exp\left[ -(t/\tau)^\beta\right]$$
where $\rho_{\infty}$ and $\rho_0$  correspond respectively to the
steady state and to the initial packing fraction value (figure~\ref{fig:compa}).
The adjustable parameters
$\tau$ and $\beta$  are here respectively 
 the relaxation time and a parameter related to 
 the stretching of the exponential.
This characteristic time scale is found to be well described 
by an Arrhenius behavior
\begin{equation}
\tau = \tau_A \exp\left[\frac{\Gamma_A}{\Gamma}\right].
\label{eqn:tau}
\end{equation}
Such kind of relaxation law is also 
found for strong glasses (the dimensionless acceleration
$\Gamma$ plays the role of the temperature).
Note that,  Lumay and Vandewalle recently recover KWW-like laws for compaction
of 2D granular packings~\cite{Lumay2005}.
As can be seen in figure~\ref{fig:compa}, a stationary state is
reached (if the number of taps is large enough).
Figure~\ref{fig:compa_finale}
reports the evolution of the packing fraction of the 
stationary state as a function of the tapping intensity
$\Gamma$ for $f=30\mbox{ Hz}$.
We observe a decrease of the
packing fraction with $\Gamma$. 
To reach an important value of packing fraction it is thus essential to use a small tapping intensity.
\section{Influence on the initial state}\label{sec:compaction_decompaction}
The question of the existence of a stationary state during 
granular compaction is essential to validate any thermodynamic approach.
As reported in the previous section, loose packings submitted 
to gentle mechanical taps continuously compact until they reach a
stationary state. Does this state depend on the initial conditions ?
To answer this question, we submit two granular packings of different
packing fractions (one dense with a packing fraction close to that
of the random close packing  : packing $1$ and one loose 
with a packing fraction
close to that of the random loose packing : packing 2) to taps of the
same intensity. 
\begin{figure}[htbp]
\includegraphics*[width=0.5\textwidth]{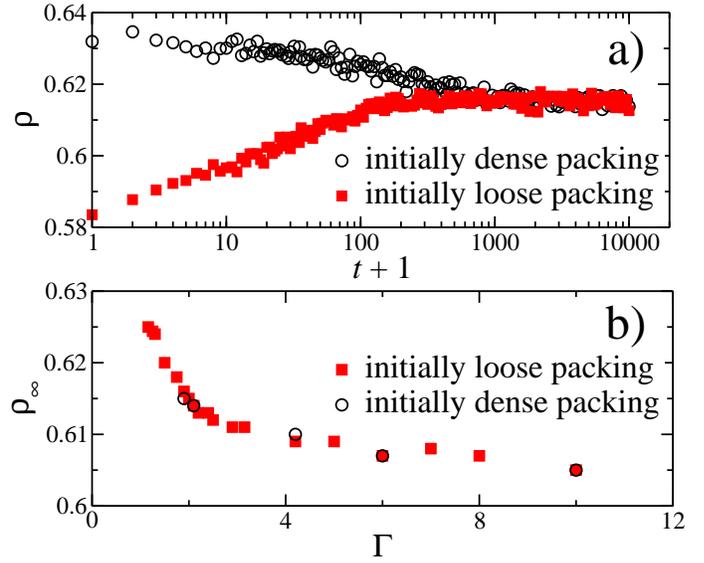}
\caption{a) Evolution of the packing fraction versus the number of taps for an initially dense packing and an initially loose packing. We observe that the final packing fraction value is similar for the two packings
($\Gamma=2.1$ and $f=30\mbox{ Hz}$).
b) Evolution of the final packing fraction versus the tapping intensity for 
an initially dense packing and an initially loose packing.}\label{fig:compaction_decompaction}
\end{figure}
As reported in
figure~\ref{fig:compaction_decompaction}a (see also~\cite{PhilippePhD2002})
after a transient the two packings reach the same value of packing fraction.
This behavior is observed for a large range of $\Gamma$
(figure~\ref{fig:compaction_decompaction}b).
This shows that, for a given tapping process, the value of the packing fraction at
stationary state does not depend on the initial conditions. 
For the two initial
packings considered, the number of taps needed to reach the steady
state is larger for decompaction than for compaction. 
This is true for any
value of $\Gamma$~\cite{PhilippePhD2002}. This can be intuitively understood : 
contrary to decompaction, compaction is facilitated by the gravity.
\\
\section{Influence of the tap duration}\label{sec:shape_tap}
\begin{center}
\begin{figure}[htbp]
\includegraphics*[width=0.5\textwidth]{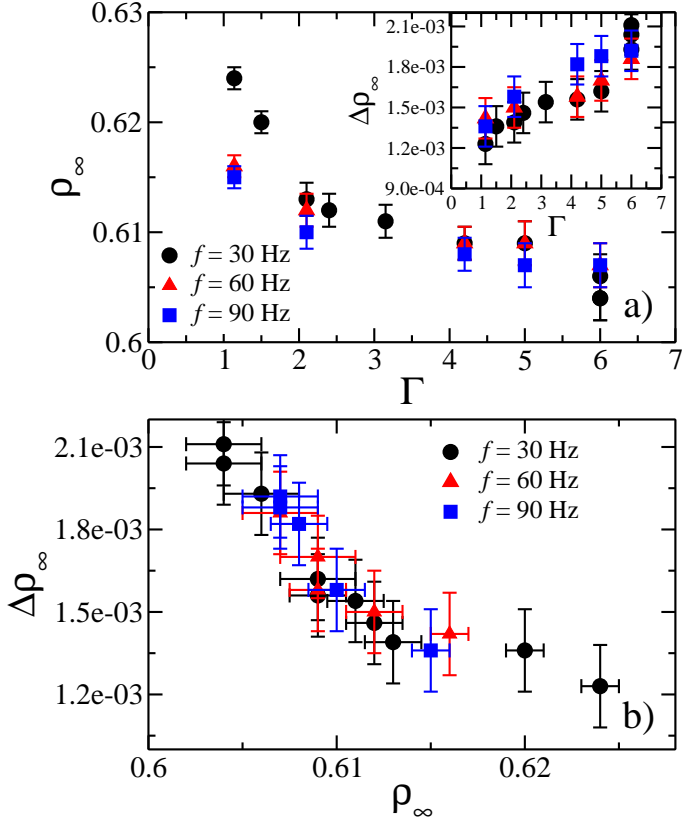}
\caption{a) The main panel shows the 
packing fraction obtained at stationarity as a function
of the tapping acceleration for different values of the frequency of the
mechanical tapping. The inset shows the  dependence of the packing fraction fluctuations 
on $\Gamma$ and $f$.
b) Plot of the packing fraction fluctuations versus 
the packing fraction at stationarity. All the data collapse on
a single master curve, showing that $\Delta\rho_\infty$
and $\rho_\infty$ are linked and that this link does not depend on
the values of $f$ and $\Gamma$.}\label{fig:fluctuation}
\end{figure}
\end{center}

\begin{figure}[htbp]
\includegraphics*[width=0.5\textwidth]{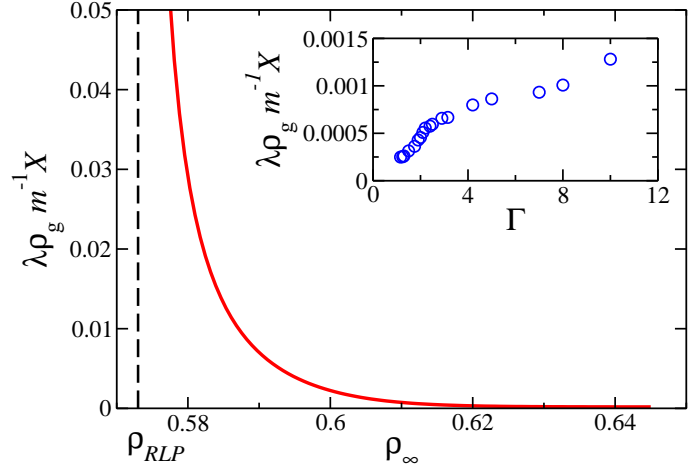}
\caption{Compactivity as a function of the average packing fraction
at stationary state. (Inset) Evolution of the compactivity versus
the tapping intensity for $f=30\mbox{ Hz}$.}\label{fig:compactivity}
\end{figure}

As mentioned in previous section, the 
stationary state does not depend on the initial
conditions. We can modify the frequency generating the
tap and study its effect on the stationary state.
We  report in Fig~\ref{fig:fluctuation}a
the stationary values of the packing fraction 
$\rho_\infty$ as well as its fluctuations $\Delta\rho_\infty$
recorded after a sequence of taps of duration $1/f$ (where
$f$ is $30$ Hz, 60 Hz or 90 Hz) and intensity 
$\Gamma$. 
Even though $\rho_\infty$ and its fluctuations
at stationary state depend more or less on both
$\Gamma$ and $1/f$ (for the latter it is particularly visible at
low values of $\Gamma$),  when $\Delta\rho_\infty$ 
is plotted as a function of $\rho_\infty$ the data collapse,
within errors bars, onto a single master curve (figure~\ref{fig:fluctuation}b). 
This shows 
that $\rho_\infty$ and $\Delta\rho_\infty$ are linked.
Note that the smallest values of $\Delta\rho_\infty$ might be
perturbated by the fluctuations of the $\gamma$-ray source.
In other words, no matter how the state with packing fraction $\rho_\infty$
is attained. This result is in agreement with those obtained very
recently by Pica Ciamarra et al.~\cite{PicaCiamarra2006} with numerical
simulations of granular packing submitted to fluid flow pulses. 
These results strongly support the idea that such stationary state are indeed genuine
"thermodynamic states". 
Moreover, in~\cite{PicaCiamarra2006}, the authors
demonstrate that at stationarity  a granular packing can be described 
by only one parameter, the packing fraction, all the other observables
being characterized by this parameter.
The strong resemblance of our results with the ones
reported in~\cite{PicaCiamarra2006} suggests that this
is also true for experiments of granular packings submitted
to mechanical tapping.\\ 
It should be pointed out that this is true only at the stationary state.
Indeed, during the transient, memory effects can be observed~\cite{Josserand2000}.
In this case, the packing fraction is in general no more enough to
fully
characterize the packing and 
 other observables are needed~\cite{Richard2006}.
Knowing the dependency of $\Delta \rho_\infty$ on $\rho_\infty$
enables us to determine Edwards' compactivity, which the equivalent
of the temperature in thermal systems. Using the fluctuation
dissipation theorem~\cite{Nowak1998} we can derive
$$\frac{\lambda\rho_g}{m}X(\rho)=\left(\int_{\rho_{{RLP}}}^{\rho}
\left(\frac{\varphi}{\Delta\varphi}\right)^2d\varphi  \right)^{-1}$$
where  we have used $X(\rho_{{RLP}})=\infty$. In this expression
$m$ is the grain mass and $\rho_g$ the grain density.
We analytically solve this equation using for $\Delta\rho_\infty(\rho_\infty)$
a 
$2^{nd}$ order polynomial
fit (figure~\ref{fig:compactivity}). In order to compare our results with those
reported in~\cite{Schroter2005} we use the same value
for the packing fraction of the random loose packing : $\rho_{{RLP}}=0.573$.
We observe a decrease of $X$ with $\rho_\infty$. 
although the evolution of 
$\Delta\rho_\infty$ is different, the behavior of $X(\rho_\infty)$  
is similar to that found for granular packings submitted to fluid
flow tapping~\cite{Schroter2005}.
We also report in the inset of figure~\ref{fig:compactivity}
the evolution of the the compactivity versus the tapping
intensity $\Gamma$ for $f=30\mbox{ Hz}$. We observed that these two quantities
are linked: the temperature-like compactivity increases with the tapping
intensity. 
Note that the change of behavior observed around $\Gamma=2$ 
may be explained by the two different convective regimes observed 
in our system~\cite{Philippe2003}.
In the analogy between glassy systems
and granular packings undergoing compaction, $\Gamma$
is supposed to play the role of temperature
at the stationary state. So the above-mentioned result reinforce
this analogy.
It should be interesting to compare the compactivity with other
``temperatures'' defined for granular media as
the granular temperature (defined as the velocity fluctuations)
or the effective temperature defined through the out-of-equilibrium
fluctuation-dissipation theorem~\cite{Song2005}.
\section{Annealing and memory effects}\label{sec:aging}
\begin{figure}[htbp]
\begin{center}
\includegraphics*[width=0.5\textwidth]{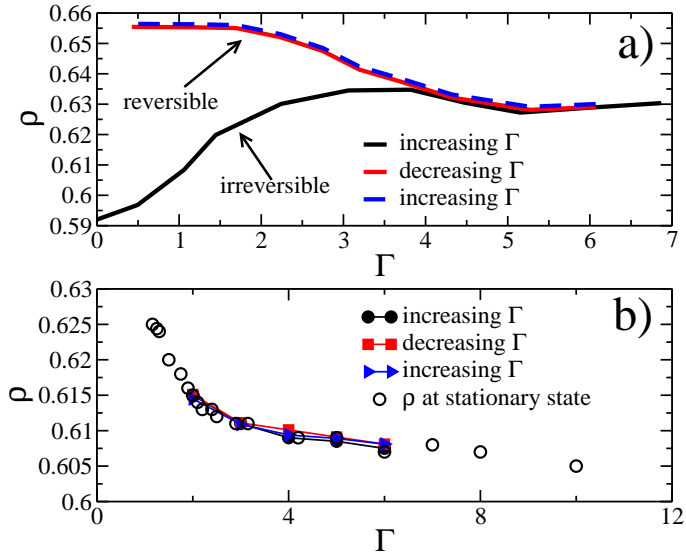}
\caption{a) Sketch of the evolution of the packing fraction during
annealing experiments reported in~\cite{Nowak1997a}.
b) Results of annealing experiments in our experiments
with $f=30\mbox{ Hz}$. No aging is observed.}\label{fig:NJ}
\end{center}
\end{figure}

Further insights into the understanding of the
stationary state characteristics can be gained by allowing the tap
intensity to vary in time.
Annealing experiments phenomena in granular
media undergoing compaction have been reported
previously in~\cite{Nowak1997a}. The authors proceeded as follows:
starting from a loose packing of grains, the material is tapped at
a given intensity $\Gamma$ for a given time $t$ ($10^5$ taps).
$\Gamma$ is then modified and the compaction process continued for $t$
taps (see figure~\ref{fig:NJ}a). The increase of $\Gamma$
corresponds to an increase of the average packing fraction
except for values larger than three for which a slow decrease can
be observed. 
If $\Gamma$ is then reduced, the packing fraction, rather than 
following reversely the previous curve, still increases. This
new curve, contrary to the previous is reversible.
Using our setup and the same protocol (with $f=30\mbox{ Hz}$), we 
only recover the reversible branch (see figure~\ref{fig:NJ}b).
This can be explained by the fact that the number of taps used
is large enough to allow our system to reach stationarity. 
  Thus, aging and irreversible-reversible
behaviors are observed only when the stationary state is not reached.
Note that these experiments also confirm that the stationary state only
depends on the tapping protocol and not on the initial conditions.\\

We also carried out memory effects experiments. Let us recall that
Josserand et al.~\cite{Josserand2000} carried out experiments
on the response function of a granular media undergoing compaction
to a sudden perturbation of the tapping intensity. These authors drive 
a granular packing to the same packing fraction
$\rho_0$ with three different tapping intensities. Then 
those three packings
are tapped at the same intensity $\Gamma_0$ and their behavior depends,
for short time, on
the previous value of the tapping intensity. 
These data show a short-time memory effect; the future evolution of the
packing fraction depends not only on its initial value but also on
its history.
Using our experimental setup we recover those memory effects as reported 
in~\cite{RibierePHD,Richard2006}.
Do such memory effects also exist at the stationary state ?
We can carry out two compaction experiments until the stationary
state is reached. For the two experiments we use two different
tapping characteristics (i.e. frequency $f_1$ and tapping
amplitude $\Gamma_1$ for experiment 1 and 
frequency $f_2$ and tapping
amplitude $\Gamma_2$ for experiment)
in such a way that the packing fractions of the two stationary
states are equal.
Once the stationary state is reached in experiment 1 we can
study the response of that packing to taps characterized by
$\Gamma_2$ and $f_2$ and compare with the behavior of the
packing at stationary state in experiment 2.
Since the history of the two packings are different one may expect
a transient in the behavior of the packing 1, sign of memory effects.
Using this
protocol 
no packing fraction modification
can be seen.
So no memory effects exist for packings at stationary state.
The range of parameters explored remains small but this preliminary
result confirms that the packing fraction is an observable that fully
characterize the stationary state.  Once more, this not true during transient
when memory effects~\cite{Josserand2000} reflect the need for, at least, an 
extra-observable to fully characterize the packing.
\section{Conclusions}\label{sec:conclu}
To Summarize, in this paper
we show that granular
packings submitted to gentle mechanical taps can reach a stationary configuration. 
This state does not depend on the initial conditions.
A dependence on the duration of the taps is found but 
our result show a one-to-one correspondence
between the final packing fraction and the packing fraction fluctuations.
The compactivity, a temperature-like state variable, is then determinated 
as a function of packing fraction of the stationary
state and as a function of the tapping intensity at a given $f$.
We also show that neither memory effects nor annealing are
present at stationary state.
Our results strongly support the relevance of a fundamental theory of
dense granular media.\\

Acknowledgments : We thank M. Nicodemi, M. Pica Ciamarra and A. Coniglio
for useful discussions and N. Taberlet for a careful reading of the manuscript.
 We acknowledge funding from French
ministry of education and research (ACI \'energie et conception 
durable ECD035 - Verres de grains) and from CNRS (PICS 3058 - Relaxation lente dans 
les milieux granulaires).



\end{document}